\documentclass{elsart}

\usepackage{harvard}
\usepackage{epsfig}

\usepackage{amssymb}

\def\url#1{{\ttfamily\def\/{/\discretionary{}{}{}}#1}}

\begin{document}

\begin{frontmatter}
\title{Dark matter dominance at all radii in the superthin galaxy UGC 7321}
\author{Arunima Banerjee\thanksref{email1}}
\address{Department of Physics,Indian Institute of Science, Bangalore 560012, India}
\author{Lynn D. Matthews\thanksref{email2}}
\address{Harvard-Smithsonian Center for Astrophysics, 60 Garden Street, MS-42, Cambridge, MA 02138, USA}
\author{Chanda J. Jog\thanksref{email3}}
\address{Department of Physics,Indian Institute of Science, Bangalore 560012, India}
\thanks[email1]{arunima$\_$banerjee@physics.iisc.ernet.in}
\thanks[email2]{Present address: MIT Haystack Observatory, Off Route 40, Westford, MA, USA 01886; email: 
lmatthew@haystack.mit.edu}
\thanks[email3]{cjjog@physics.iisc.ernet.in}

\begin{abstract}
We model the shape and density profile of the dark matter halo of the low surface brightness, superthin galaxy UGC 7321, using the observed rotation curve and the {\mbox{H\,{\sc i}}} scale height data as simultaneous constraints. We treat the galaxy as a  gravitationally coupled system of stars and gas, responding to the gravitational potential of the dark matter halo. An isothermal halo of spherical shape with a core density in the range of 0.039 - 0.057 $M$$_{\odot}$ $pc$$^{-3}$ and a 
core radius between 2.5 - 2.9 $kpc$, gives the best fit to the observations for a range of realistic gas parameters assumed. We find that the best-fit core radius is only slightly higher than the stellar disc scale length (2.1 $kpc$), unlike the case of the high surface brightness galaxies where the halo core radius is typically 3-4 times the disc scale length of the stars. Thus our model shows that the dark matter halo dominates the dynamics of the low surface brightness, superthin galaxy 
UGC 7321 at all radii, including the inner parts of the galaxy. \\
\end{abstract}

\begin{keyword}

galaxies: ISM  \sep galaxies: kinematics and dynamics  
\sep galaxies: spiral \sep galaxies: structure 
\sep galaxies: halos \sep galaxies: individual: UGC 7321

\end{keyword}
\end{frontmatter}

\section{Introduction}
\label{sec:intro}

Since spiral galaxies are rotationally supported systems, disc rotation curves generally serve as valuable tracers of the gravitational potential in the galactic plane. Through traditional mass-modelling, the observed curve is routinely used to infer the mass distribution of galaxies and hence their dark matter contents (e.g. Begeman 1987; Kent 1987; Geehan et al. 2006). In contrast, the thickness of the gas layer depends on the vertical gravitational force and thus traces the potential perpendicular to the plane e.g.,(Narayan \& Jog 2002a).  

Recently, the rotation curve and the outer galactic {\mbox{H\,{\sc i}}} flaring data have been used together to probe the dark matter halos of a few galaxies. The rotation curve mainly determines the mass enclosed within a given radius, and therefore the power-law index of the density profile of the halo. The flaring curve, on the other hand, determines its shape uniquely. So, both the constraints have to be used on an equal footing to correctly determine the parameters of the dark matter halo of any galaxy.

The {\mbox{H\,{\sc i}}} scale height data coupled with the rotation curve has been used to study the dark matter halos of NGC 4244 (Olling 1996) , NGC 891 (Becquaert \& Combes 1997) and the Galaxy (Olling \& Merrifield 2000, 2001) in the past. Narayan et al. (2005) studied the Galactic dark matter halo by rigorously incorporating the self-gravity of the gas into their model for the Galaxy unlike some of the previous studies given in the literature. They concluded that a steeper-than-isothermal, spherical halo best fits the observations, the scale height data at that time being available up to galactocentric distances of 24 $kpc$. These results were confirmed by Kalberla et al. (2007), who, however, included a dark matter ring in their model to explain their extended {\mbox{H\,{\sc i}}} scale height data available till 40 $kpc$. In our previous work (Banerjee \& Jog 2008), we studied the dark matter halo of M31, where we developed a model similar to the Galaxy (Narayan et al. 2005) . However, in addition, we included the bulge into the model, and also varied the shape of the halo as a free parameter, unlike the Galaxy case. Further, we fitted the rotation curve over the entire radial range instead of pinning it at a single point like the Galaxy case. We scanned the four dimensional grid of the four free parameters characterizing the halo, in a systematic manner, and found that an isothermal halo of an oblate shape of axis ratio $q$ = 0.4 gives the best fit to the available data. 

In this paper, we apply for the first time a similar approach to study the dark matter halo properties of a low surface brightness (LSB) ``superthin''  galaxy: UGC 7321. UGC 7321 is a bulgeless, pure disc galaxy of Hubble Type Sd, and has a highly flattened stellar disc with a planar-to-vertical axis ratio of 10.3. A few of its key properties are summarized in Table 1. The galaxy has an extended {\mbox{H\,{\sc i}}} disc, and the scale height data are available up to 6-7 disc scale lengths (Matthews \& Wood 2003). So it is highly suitable for the application of the above method to probe its dark matter halo properties.

\begin{table*}
 \begin{minipage}{140mm}
\caption{UGC 7321 Paramaters}\footnote{All quantities are taken from Matthews et al. (1999) and Uson \& Matthews (2003) which assume d = 10 $Mpc$}
\label{tab:gmrt}
\vskip 0.1in
\begin{tabular}{ll}
\hline
Parameters& Value \\
\hline
\hline
$A_{opt}$($kpc$) & $16.3$ \footnote{Linear diameter at limiting observed $B$-band isophote of 25.5 $mag$ $arcsec^{-2}$} \\
$L_{B}$($L_{\odot}$) & $1.0$ $\times$ $10^{9}$ \footnote{Blue luminosity} \\
$M_{{\mbox{H\,{\sc i}}}}$($M_{\odot}$) & $1.1$ $\times$ $10^{9}$ \footnote{${\mbox{H\,{\sc i}}}$ mass}\\
$h_{R}$($kpc$) & $2.1$ \footnote{disc scale length measured from $R$-band image}\\
$z_{0}$($pc$) & $150$ \footnote{stellar scale height obtained from an exponential fit}\\
$\mu_{B,i}(0)$($mag$ $arcsec^{-2}$) & $23.5$ \footnote{Deprojected (face-on) central disc surface
   brightness in the $B$ band, corrected for internal and Galactic
   extinction}  \\ 
$v_{rot}$($km$$s^{-1}$) & $105$ \\
$Star$ $formation$ $rate$ ($M_{\odot}$ per year for massive stars $\ge 5 M_{\odot}$ & $\sim 0.02$ \\
\hline
\end{tabular}
\end{minipage}
\end{table*}

Based on traditional mass-modelling which only uses the observed rotation curve as the constraint, it has been found that the late-type, low surface brightness galaxies are generally dark matter dominated, often within the inner portions of their stellar discs (de Blok \& McGaugh 1997) . In the case of UGC 7321, other lines of 
evidence have already suggested that it, too, is a highly dark 
matter-dominated galaxy. It has large ratios of its dynamical mass 
to its {\mbox{H\,{\sc i}}} mass and blue luminosity, ($M_{\rm dyn}/M_{HI}$ = 31 and $M_{\rm dyn}/L_{B}$ = 29, respectively; (cf. Roberts \& Haynes 1994) , and an extraordinarily small stellar disc scale height 
($\sim$150~$pc$ for a distance of 10~$Mpc$ based on an exponential fit; Matthews 2000). These properties suggest the need for a massive dark halo to stabilize the disc 
against vertical bending instabilities (Zasov et al. 1991) .

UGC 7321 is devoid of a central bulge component (Matthews et al. 1999)
and its molecular gas content appears to be dynamically insignificant
(Matthews \& Gao 2001; Matthews \& Wood 2001). We, therefore, model the
galaxy as a gravitationally coupled, two-component system of stars and atomic hydrogen gas with the dark matter halo acting as a source of external force to this system. We use a four-parameter density profile for the dark matter halo (de
Zeeuw \& Pfenniger 1988; Bequaert \& Combes 1997): the core density, the core radius, the power-law density index and the axis ratio of the halo being the four free parameters characterizing it. We methodically vary the four parameters within their respective feasible ranges, and try to obtain an optimum fit to both the observed rotation curve and the vertical scale height data at the same time. As we shall see, this method predicts a spherical, isothermal halo with a core density of about 0.039- 0.057 $M_{\odot}$ $pc$$^{-3}$ and core radius of 2.5 - 2.9 $kpc$ for this galaxy. 

The layout of the present paper is as follows. We briefly discuss the model in \S 2, and in \S 3 the method of solving the equations and the input parameters used is discussed. In \S 4, we present the results, followed by the discussion and conclusions in \S\S 5 and 6, respectively.

\section{Description of the model used}
\label{sec:des}
\subsection{Gravitationally coupled, two-component, galactic disc model}
\label{ssec:grav_coup}

The galaxy is modelled as a gravitationally-coupled, two-component system of stars and atomic hydrogen gas embedded in the dark matter halo, which exerts an external force on the system  while remaining rigid and non-responsive itself. This is a simplified version of the Galaxy case (Narayan \& Jog 2002b) , where a gravitationally-coupled, three-component system of stars, atomic and molecular hydrogen was considered. Here, the two components, present in the form of discs, are assumed to be axisymmetric and coplanar with each other for the sake of simplicity. Also, it is assumed that the components are in a hydrostatic equilibrium in the vertical direction. Therefore, the density distribution of each component will be jointly determined by the Poisson equation, and the corresponding equation for pressure equilibrium perpendicular to the midplane. 

In terms of the galactic cylindrical co-ordinates ($R$, $\phi$, $z$), the Poisson equation for an azimuthally symmetric  system  is given by \\
$$\frac{{\partial}^2{\Phi}_{total}}{{\partial}z^2} + \frac{1}{R}\frac{{\partial}}{{\partial}R}(R \frac{{\partial}\Phi_{total}}{{\partial}R})
 = 4\pi G(\sum_{i=1}^{2} \rho_i + \rho_{h})
\eqno(1) $$ \\
where $\rho_i$ with i = 1 to 2 denotes the mass density for each disc component while $\rho_h$ denotes the mass density of the halo. $\Phi_{total}$ denotes the total potential due to the disc and the halo. For a nearly constant rotation curve as is the case here, the radial term can be neglected as its contribution to the determination of the {\mbox{H\,{\sc i}}} scale height is less than ten percent as was noted by earlier calculations Narayan et al. (2005). So, the above equation reduces to \\
$$\frac{{\partial}^2\Phi_{total}}{{\partial}z^2}
 = 4\pi G(\sum_{i=1}^{2} \rho_i + \rho_{h})
\eqno(2) $$ 
The equation for hydrostatic equilibrium in the z direction is given by Rohlfs (1977)
$$ \frac{\partial}{{\partial}z}(\rho_{i}\langle(v_{z}^{2})_{i}\rangle) + \rho_{i}\frac{{\partial}\Phi_{total}}{{\partial}z} = 0  \eqno(3) $$ \\
\noindent where $\langle(v_{z}^{2})_{i}\rangle$ is the mean square random velocity along the $z$ direction for the $i^{th}$ component. Further we assume that each component is isothermal i.e., the random velocity $v_{z}$ remains constant with $z$. 

Combining eq. (2) and eq. (3),  we get
$$ \langle(v_{z}^{2})_{i}\rangle  \frac{\partial}{{\partial}z}[\frac{1}{\rho_{i}}\frac{{\partial}\rho_{i}}{{\partial}z}] = -4\pi G(\sum_{i=1}^{2} \rho_i + \rho_{h})
 \eqno(4)$$

\noindent This represents a set of two coupled, second-order, ordinary differential equations which needs to be solved to obtain the vertical density distribution of each of the two components. Although the net gravitational potential acting on each component is the same, the response will be different due to the different velocity dispersions of the two components.

\subsection{ Dark Matter Halo }
\label{ssec:DM halo}

We use the four-parameter dark matter halo model (de
Zeeuw \& Pfenniger 1988; Bequaert \& Combes 1997) with the density profile given by

$$\rho(R,z) = \frac{\rho_0}{\large [ 1+\frac{m^{2}}{{{R_c}}^{2}}\large]^p} \eqno(5) $$  \\
where $ m^{2}$=$R^{2} + ({z^{2}}/{q^{2}})$, 
 $\rho_0$  is the central core density of the halo, ${R_c}$
is the core radius, $p$ is the power-law density index, and $q$ is the vertical-to-planar axis ratio of the halo 
(spherical: $q$ = 1; oblate: $q$ $<$ 1; prolate: $q$ $>$ 1).

\section {Numerical Solution of the Equations \& Input Parameters}
\label{sec: Num}
\subsection{Solution of equations}
\label{ssec: Sol}

For a given halo density profile, eq. (4) is solved in an iterative fashion, as an initial value problem, using the fourth-order, Runge-Kutta method of integration, with the following two initial conditions at the
mid-plane (i.e., $z$ = 0) for each component:
$$ \rho_i = (\rho_0)_i,  \qquad \frac{d\rho_i}{dz} = 0  \eqno(6) $$ \\
As the modified mid-plane density $(\rho_0)_i $ for each
component is not known a priori, the net surface
density $\Sigma_i(R)$,
given by twice the area under the curve of
$\rho_i(z)$ versus z, is used as the secondary boundary condition,
as this quantity is known from observations (see \S 3.2).
The required value of $(\rho_i)_0$
is thus determined by a trial and error method, which
gives the required $\rho_i(z)$ distribution after four iterations with an accuracy to the second decimal place. Existing theoretical models suggest a sech$^2$ profile for an isothermal density distribution. But for a three-component disc, the vertical distribution is shown to be steeper than a sech$^2$ function close to the mid-plane (Banerjee \& Jog 2007).However, at large $z$ values, it is close to a sech$^2$ distribution. Hence we use the half-width-at-half-maximum of the resulting model vertical distribution to define the scale height as was done in Narayan \& Jog (2002a,b).

\subsection {Input Parameters}
\label{ssec: Input}

We require the vertical velocity dispersion and the surface density of each of the two galactic disc components to solve the coupled set of equations at a given radius. The central stellar surface density is derived directly from the optical surface photometry (Matthews et al. 1999) by assuming a reasonable stellar mass-to-light ratio. The deprojected $B$-band central surface brightness of UGC~7321 (corrected for extinction) translates to a central luminosity density of
26.4~$M_{\odot}$~pc$^{-2}$. Using the $B-R$ color of the central regions
($\sim$1.2; Matthews et al. (1999) ) and the ``formation epoch: bursts'' models
from Bell \& de Jong (2001) predicts $(M/L)_{\star}$ = 1.9, which we adopt
here. (Other models by Bell \& de Jong give values of $(M/L)_{\star}$ ranging
from 1.7 to 2.1). This in turn yields a central stellar surface density of
50.2~$M_{\odot}$ pc$^{-2}$ for UGC~7321.

The stellar velocity dispersion of this galaxy has been indirectly estimated to be 14.3 $km$$s^{-1}$ at the centre of the 
galaxy ($R$ = 0) (Matthews 2000). This is very close to the value of the central (vertical) stellar velocity dispersion (16 $km$$s^{-1}$) for the dwarf spiral galaxy UGC 4325 measured by Swaters (1999), and to the value (20 $km$$s^{-1}$) estimated analytically for the superthin galaxy IC 5249 by van der Kruit et al. (2001). We assume the central value of velocity dispersion to fall off exponentially with radius  with a scale length of 2 $R_{d}$ (which is equal to 4.2 $kpc$ for UGC 7321) as is seen in the Galaxy (Lewis \& Freeman 1989).
Uson \& Matthews (2003) give the deprojected {\mbox{H\,{\sc i}}} surface density for UGC 7321 as a function of radius. The velocity dispersion of {\mbox{H\,{\sc i}}} is obtained from the Gaussian fits to the edges of position-velocity cuts on the observed data. This gives a value between 7-9 $km$$s^{-1}$. The data are consistent to the typical value of the {\mbox{H\,{\sc i}}} dispersion in other galaxies (See \S 5.2 for a detailed discussion).

The molecular hydrogen gas, H$_{2}$, has not been taken into account, as 
it appears to be dynamically insignificant compared to the other 
components of the disc. Matthews
\& Gao (2001) detected a weak CO signal 
from the central $\sim$2.7~$kpc$ of UGC~7321, which translates to a total 
molecular hydrogen mass of H$_{2}\approx2.3\times10^{7}~M_{\odot}$ 
(although this value is uncertain by at least a factor of 2-3 as a result 
of uncertainties in optical depth effects and the appropriate value of the 
CO-to-H$_{2}$ conversion factor). This corresponds to a mean H$_{2}$ 
surface density of $\Sigma_{H2}\approx 1~M_{\odot}$ in the inner galaxy, 
which agrees fairly well with independent estimates from the dust models 
of Matthews \& Wood (2001) and from a study of the distribution of dark 
clouds from {\it Hubble Space Telescope} images (J. S. Gallagher \& L. D. 
Matthews, unpublished). Therefore, the presence of H$_{2}$ has been 
ignored in subsequent calculations.

\section{Results and analysis}
\label{sec: Results}

We perform an exhaustive scanning of the grid of parameters characterizing the dark matter density profile to obtain an 
optimum fit to both the observed rotation curve and the scale height data. To start with, we consider a spherical halo ($q$ = 1) for simplicity. 

\begin{table*}
 \begin{minipage}{140mm}
\caption{3D grid of dark halo parameters scanned}
\label{tab:gmrt}
\vskip 0.1in
\begin{tabular}{lll}
\hline
$Parameter$ & $Range$ & $Resolution$\\
\hline
$\rho_{0}$($M$$_{\odot}$ $p$c$^{-3}$) & $0.0001 - 0.1$ & $0.0001$\\
				      & $0.001 - 0.5$  & $0.001$  \\
$R_{c}$($kpc$) & $1.5 - 12$ & $0.1$ \\
$p$ & $1 - 2$ & $0.5$ \\
\hline
\hline
\end{tabular}
\end{minipage}
\end{table*}

We vary the remaining three free parameters characterizing the density profile of the halo (see eq. (5)) within their respective feasible ranges (as summarized in Table 2), and obtain the contribution of the halo to the rotation curve for each such grid point in this three-dimensional grid. The power-law density index $p$ is allowed to take the values 1, 1.5 and 2 successively. Here, a value of $p$ = 1 corresponds to the standard isothermal case used routinely for simplicity and also because it corresponds to the flat rotation curve. The value of $p$ = 1.5 refers to the NFW profile Navarro et al. (1996) at large radii, whereas $p$ = 2 gives an even steeper dark-matter halo profile, as was found for the Galaxy case Narayan et al. (2005). For each value of $p$, the core density $\rho_{0}$ and the core radius $R_{c}$ are varied as given in Table 2 to ensure an exhaustive scanning for the dark matter halo parameters since we have little prior knowledge of the plausible values these parameters can take in a superthin galaxy. 

\subsection{The rotation curve constraint}
\label{sec: rotcurve}

\noindent The total rotational velocity at each radius is obtained by adding the contribution from the stars, the gas and halo in quadrature as

$$ {v^{2}(R)} = v_{star}^{2}(R) + v_{gas}^{2}(R) + v_{halo}^{2}(R) \eqno(7) $$ 	\\
Here the way to obtain the different terms is discussed below. This result is matched with the observed rotational velocity 
at all radii.

The deprojected gas surface density versus radius data for UGC 7321 (Uson \& Matthews 2003) can be modelled as one which remains constant at 5 $M$$_{\odot}$ $pc$$^{-2}$ at galactocentric radii less than 4 $kpc$, and which then falls off exponentially with a scale length of 2.8 $kpc$. The gas surface density does not include a correction for He. For this radial distribution, we calculated the contribution of the gas to the rotation curve (using eq. (2-158) \& (2-160) of Binney \& Tremaine (1987)), and found it to be negligible compared to that of the stellar component.  However, it was included in the calculations for the sake of completeness. 

The rotational velocity at any radius $R$ for a thin exponential stellar disc is given by Binney \& Tremaine (1987)

$$ v_{star}^{2}(R) = 4\pi G \Sigma_{0} R_{d} y^{2} [I_0(y)K_0(y) - I_1(y)K_1(y)] \eqno(8)$$ 	\\
where $\Sigma_{0}$ is the disc central surface density, $R_{d}$ the disc scale length and $y$ = $R$/{2$R$$_{d}$}, $R$ being the galactocentric radius. The functions $I_{n}$ and $K_{n}$ (where n=0 and 1) are the modified Bessel functions of the first and second kind, respectively.

For the spherical halo, the rotational velocity, $v_{halo}(R)$, is given by \\
$$ v_{halo}^{2}(R) = \frac{G M_{halo} (R)}{R} \eqno(9) $$ \\
where  $M$$_{halo} (R)$, the mass enclosed within a sphere of radius $R$ for a the given halo density profile, and is obtained from the density as given by the right-hand side of eq. (5).

 For an oblate halo of axis ratio $q$ and density index $p$, the circular speed $v_{halo}(R)$ is obtained by differentiating the expression for the potential from Sackett \& Sparke (1990), and Becquaert \& Combes (1997) to be: \\
$$ v_{halo}^{2}(R) = 4 \pi G \rho_{0} q \int_{0}^{1/q} \frac{R^2 x^2 [ 1 + \frac{R^2 x^2}{R_{c}^2 ( 1 + \epsilon^2 x^2)}  ]^{-p}}{( 1 + \epsilon^2 x^2)^2} dx						\eqno(10)$$ \\
\noindent where $\epsilon = (1- q^2)^{1/2}$.
We obtain the value of the integral numerically in each case.

Thus upon obtaining the rotation curve corresponding to each grid point, we perform the ${\chi}^{2}$ analysis comparing computed to the observed {\mbox{H\,{\sc i}}} rotation curve. The observed rotation curve is taken from Uson \& Matthews (2003) and has 30 data points with very small error-bars (typically a few percent of the observed velocity amplitudes even after accounting for systematic uncertainties).
It was derived by implicitly assuming a constant (Gaussian) {\mbox{H\,{\sc i}}} velocity dispersion of 7 $kms^{-1}$. Ideally, we should have considered only those grid points which give ${\chi}^{2}$ values of the order of 30 (i.e., the number of data points) as those giving appreciably good fits to the observed curve Bevington (1969). But we relax this criterion and choose a larger range of grid points around the minimum i.e grid points which give ${\chi}^{2}$ values less than 300 for applying the next constraint i.e the vertical {\mbox{H\,{\sc i}}} scale height data. This allows us to impose the simulataneous constraints (planar + vertical) on our model. (See \S 4.3 for a discussion). So finally we get 36 grid points for $p$ = 1, 80 for $p$ = 1.5 and 69 for $p$ = 2 case.  As we shall see later, the final set of best-fit parameters obtained give reasonably good fits to both the observed rotation curve and the scale height data.

\subsection{The {\mbox{H\,{\sc i}}} scale height constraint}
\label{sec: HI scaleheight}

For each value of $p$, we obtain the {\mbox{H\,{\sc i}}} scale height distribution beyond 3 disc scale lengths, for each of the grid points filtered  out by the first constraint as discussed in the previous section. Next we perform the ${\chi}^{2}$ analysis of our model {\mbox{H\,{\sc i}}} scale height versus radius curves with respect to the observed one and try to fit our model to the observed data only beyond 3 disc scale lengths in keeping with the earlier studies in the literature (Narayan \& Jog 2005; Banerjee \& Jog 2008). For M31, the surface-density and therefore the vertical gravitational force due to the dark matter halo exceeds that of the disc only in the outer regions (See Fig.6 of Banerjee \& Jog 2008). As the disc dynamics in this region are controlled by the halo alone, the above method helps us in studying the effect of the halo on the scale height distribution, decoupled from that of the other components. For the case of UGC 7321, at first we take the gas velocity dispersion to be equal to 7 $km$$s^{-1}$ . However it fails to give a good fit to the observed data. Next we try both 8 $km$$s^{-1}$ and 9 $km$$s^{-1}$ successively, but choose the latter for subsequent calculations as it gives much better fit to the observed data as compared to the 8 $km$$s^{-1}$ case. 

 For the choice of $v_{z}$ = 9 $kms^{-1}$, the best-fit core density is 0.041 $M$$_{\odot}$$pc$$^{-3}$ and a core radius is 
2.9 $kpc$, as indicated by the smallest ${\chi}^{2}$ value. The small value of the best-fit halo core radius thus obtained indicates that the halo becomes important already at small radii. This suggests that the fitting of the theoretical curve with the observed one should not be restricted only to regions beyond 3 $R_{d}$ for an LSB galaxy like UGC 7321 as the halo is already important at small radii. Hence, we next fit the scale height data over entire radial range (i.e., 2-12 $kpc$) with the same constant $v_{z}$ value of 9 $kms^{-1}$. The best-fit values change by less than a few percent compared to the above case where the fit was done only beyond 3 $R_{d}$. The best-fit core density now becomes 0.039 $M$$_{\odot}$$pc$$^{-3}$ wheras the best-fit core radius continues to be 2.9 $kpc$. 

Since the disagreement of the observed rotation curve with the predicted one is mostly in the inner galaxy, we check if the fit can be improved by reducing the central value of the stellar surface density by twenty percent or so, keeping the $v_{z}$ value contant at 9 $km$$s^{-1}$. This is reasonable as there are uncertainties of at least that order in evaluating both the $M$/$L$ ratio and the deprojected surface brightness of the stellar disc. However, this variation fails to improve the results significantly.

We then take a cue from the nature of the mismatch of our model curve with the observed one, which clearly shows the need to use a higher value of gas velocity dispersion in the inner parts, while a slightly lower value is required in the outer regions. Also, the nature of the mismatch rules out an oblate halo as a possible choice as that will lower the scale heights throughout the entire radial range, thus making the fits worse in the inner regions. To account for this, we then repeat the whole procedure by imposing a small gradient in the gas velocity dispersion by letting it vary linearly between 9.5 $km$$s^{-1}$ at $R$ = 7 $kpc$ and 8 $km$$s^{-1}$ at $R$ = 12.6 $kpc$. Although such a variation is ad-hoc, the observational constraints on this value are weak enough to allow for a small variation with galactocentric radius, with 9.5 $km$$s^{-1}$ approaching the upper limit allowed by the data. Using the same gradient in the inner regions, we get a gas velocity dispersion of 10.8 $km$$s^{-1}$ at $R$ = 2 $kpc$. We may note here that a similar gradient in the {\mbox{H\,{\sc i}}} velocity dispersion was obtained in the case of the Galaxy (Narayan \& Jog 2002b) and led to a better fit to the observed scale height in the inner Galaxy (See \S 5 for a detailed discussion). A fit to the whole range of observations (2 - 12 $kpc$) gives an isothermal halo of spherical shape with a core density of 0.043 $M$$_{\odot}$ pc$^{-3}$ and a core radius of 2.6 $kpc$ best fits the observations. 
These values are only slightly different (within 10 percent) from the values obtained with a constant velocity of $9 kms^{-1} $.

In Fig.1, we give our best-fit for the case of constant $v_{z}$ = 9 $km$$s^{-1}$, and the case with a $v_{z}$ slightly falling with radius as compared to the fit to the rotation curve alone, superimposed on the observed one. Our model curves follow the trend of the observed data well throughout the entire radial range. 

\begin{figure}[h!]
\begin{center}
\rotatebox{0}{\epsfig{file=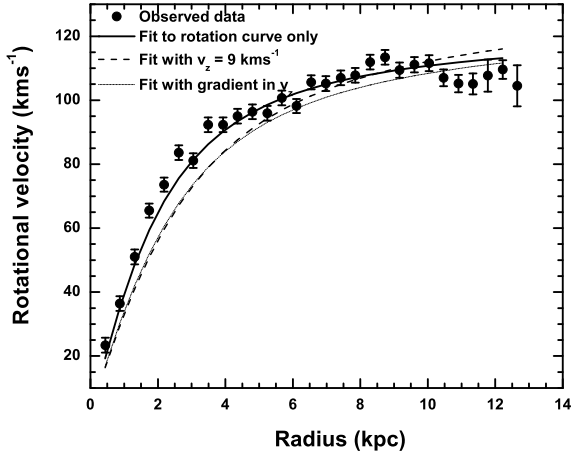,width=3.5in}}
\end{center}
\caption{Plot of the rotational velocity (in $kms^{−1}$) versus radius (in kpc) for the
best-ﬁt case of a spherical isothermal halo and a constant gas velocity dispersion
of vz = 9 $kms^{−1}$ (dashed line) and for the case of vz falling slightly with radius
(dotted line) superimposed on the best-ﬁt to the rotation curve alone. Overall,
the model rotation curves follow the trend of the observed data.
}
\label{fig:rotcur}
\end{figure}
    
In Fig.2, we compare the best-fit scale height distributions for the above two cases with the observed one. Clearly, the case with a gradient in gas velocity dispersion gives a remarkably better fit (${\chi}^{2}$ value 2.8), although as far as ${\chi}^{2}$ values are concerned, the case of constant $v_{z}$ = 9 $km$$s^{-1}$ cannot be ruled out altogether (${\chi}^{2}$ value 14.7) (This is because basic statistics suggests that the fit to the model is considered to be reasonably good if the ${\chi}^{2}$ value is of the order of the number of data points in the fit as discussed earlier at the end of \S 4.1. Here the total number of data points in the {\mbox{H\,{\sc i}}} scale height data is 11.)

\begin{figure}[h!]
\begin{center}
\rotatebox{0}{\epsfig{file=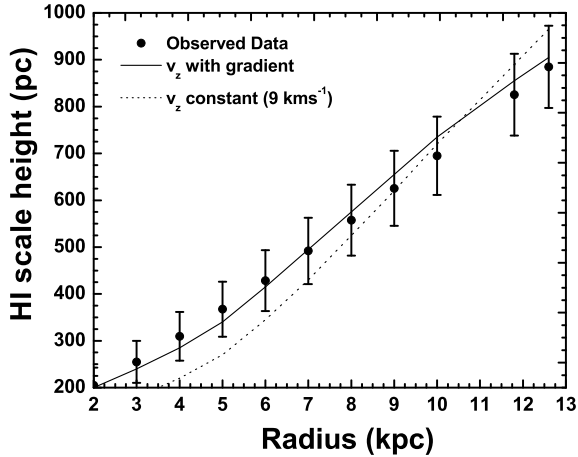,width=3.5in}}
\end{center}
\caption{Plot of {\mbox{H\,{\sc i}}} scale height (in pc) versus radius (in $kpc$) for the best-fit case of a spherical isothermal halo with constant gas velocity dispersion ($v_{z}=9$~$km$s$^{-1}$; dotted line) and for 
$v_{z}$ declining slowly with radius (solid line). In this case, the model curves have been fitted over the entire radial
range. The model with constant $v_{z}$ predicts an {\mbox{H\,{\sc i}}} scale height distribution that does not reproduce the observed values in the inner regions of the galaxy ($R<7$~$kpc$). Assuming a slight gradient in $v_{z}$ clearly gives a 
better fit}
\label{fig:scht}
\end{figure}

\subsection{Quality of individual fits as a result of imposing two simultaneous constraints}

We reiterate the fact that our method is aimed at obtaining an optimum fit to both the observational constraints, namely the rotation curve and the HI scale height data. This evidently results in a compromise in the quality of individual fits to either of the observed curves (See Fig.1 \& 2). Traditional mass modelling techniques resort to the rotation curve constraint 
alone, and therefore the fit is much better. However imposing two simultaneous constraints on the theoretical model gives a 
more realistic picture than the case in which best-fit is sought to a single constraint alone. It is noteworthy that even when the fit is sought to the rotation curve alone, the best-fit $R_{c}$ continues to be of the order of $R_{D}$ which is tha main result of this work. However the $\rho_{0}$ value obtained is different in the two cases.


\section{Discussion}
\label{sec:dis}

The dark halo properties and overall stability of superthin galaxies like UGC~7321 are of considerable interest in the
context of galaxy formation and evolution. In particular, such galaxies seem to pose a significant challenge to
hierarchical models of galaxy formation, whereby galaxies are built-up through violent mergers of subgalactic clumps
 since such mergers may result in significant disc heating and trigger instabilities (e.g., D'Onghia \& Burkert 2004, Eliche-Moral et al. 2006, Kormendy \& Fisher 2005). While theorists have predicted that the thinnest galaxy disks must require massive dark halos for stabilization (Zasov et al. 1991; Gerritsen \& de Blok 1999), little information has been available on the dark halo properties of individual superthin galaxies until now.

UGC 7321 is the first superthin galaxy for which both a detailed rotation curve and the gas layer thickness were derived
Uson \& Matthews (2003). This has allowed us to use both these constraints simultaneously to
characterize its dark halo properties, as well as to obtain new insight into the stability of its disc against star
formation. Below we comment further on the implications of several of our key findings.

\subsection {The small core radius of the dark matter halo}
\label{ssec: core radii}
  
The core radii of the dark matter halos of massive high surface brightness galaxies studied so far are usually  found to be comparable to their optical size, or equivalently, 3-4 times larger than the exponential stellar disc length Gentile et al. (2004). The Galaxy has a core radius of 8-9.5 $kpc$ which is equal 3$R_{d}$ (Narayan et al. 2005) while M31 has a core radius equal to 21 $kpc$ which is almost equal to 4$R_{d}$ (Banerjee \& Jog 2008). For UGC 7321, we find a very small core radius of 2.5-2.9 $kpc$, which is just slightly greater than its disc scale length ($R_{d}$ = 2.1 $kpc$). This shows that the dark matter becomes important at small radii consistent with previous mass-modelling of LSB spirals, based on other techniques (de Blok \& McGaugh 1997; de Blok et al. 2001). This is illustrated in another way in Fig.3, which gives a comparative plot of the surface-density of the stars, gas and the halo with radius in this galaxy. The halo surface density was calculated within the total gas scale height as was done for M31 (Banerjee \& Jog 2008). It clearly shows that the surface-density and hence the gravitational potential of the halo becomes comparable to that of the disc already at R = 2$R_{d}$. This behaviour is quite different from that of a high surface density galaxy like M31 (cf Fig.6, Banerjee \& Jog 2008), where the halo contribution starts to dominate at much larger radii (5$R_{d}$). Our results support the idea that superthin disks like UGC 7321 are among the most dark matter-dominated of disc galaxies.

\begin{figure}[h!]
\begin{center}
\rotatebox{0}{\epsfig{file=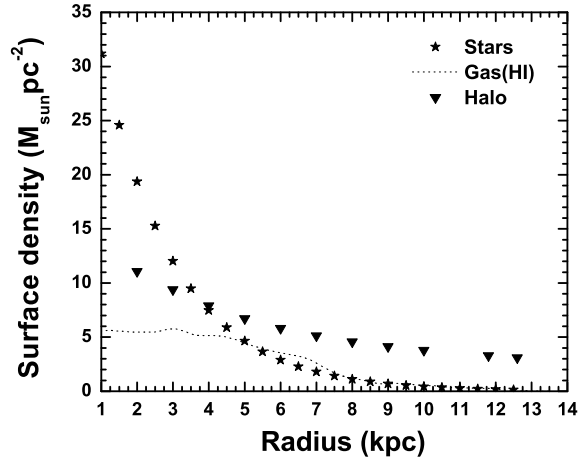,width=3.5in}}
\end{center}
\caption{Plot comparing the surface-density (in $M_{\odot}$ $pc^{-2}$) of the stars, gas and the dark-matter halo as a function of radius (in $kpc$). It clearly shows that the gravitational potential of the halo dominates over that of the disc as early as two disc scale lengths ($r$ = 4 $kpc$)}
\label{fig:sdensity}
\end{figure}

\subsection{Dependence on gas parameters}

$\bullet$ \textbf{Gradient in gas velocity dispersion} As noted earlier, if we impose a constant velocity dispersion, we require a value of 9 $km$$s^{-1}$ to get a reasonably good fit to the observed scale height data, while an even better fit requires a velocity gradient implying even larger dispersion in the inner region (Fig.2). In the earlier work for the Galaxy (Narayan et al. 2005), a slope of -0.8 $km$$s^{-1}$ $kpc^{-1}$ for the gas velocity dispersion was obtained for the inner Galaxy between 2-12 $kpc$ (pinned at 8 $km$$s^{-1}$ at 8.5 $kpc$) as it gave the best-fit to the nearly constant {\mbox{H\,{\sc i}}} scaleheights. Oort (1962) had tried the same idea but had needed a higher gradient of -2 $km$$s^{-1}$ $kpc^{-1}$ since he did not include the gas gravity and therefore needed a larger variation to account for the constant {\mbox{H\,{\sc i}}} scaleheight in the inner Galaxy. Narayan et al. (2005) tried to constrain the halo properties using the outer galaxy {\mbox{H\,{\sc i}}} data, where they had used gas velocity gradient of -0.2 $km$$s^{-1}$ $kpc^{-1}$. This is similar to the value we have for UGC 7321. This was based on the fact that some galaxies show a falling velocity dispersion which then saturates to 7 $km$$s^{-1}$ (See Narayan et al. 2005 for a discussion). Recently, Petric \& Rupen (2007) have measured the {\mbox{H\,{\sc i}}} velocity dispersion across the disc of the face-on galaxy NGC 1058. The authors find the {\mbox{H\,{\sc i}}} velocity dispersion to have a fairly complex distribution, but nonetheless show a clear fall-off with radius (see Fig.8 of their paper). Using this figure, one can estimate a gradient of roughly -0.1 $km$$s^{-1}$$kpc^{-1}$ in the outer disc, which is consistent with values observed for other galaxies. A similar fall-off has also been seen in NGC 6946 (Boomsma et al. 2008) as well as in several other galaxies (Bottema et al. 1986; Dickey et al. 1990; Kamphius 1993). So this gives some observational support to our assumption.\\ 
 
$\bullet$ \textbf{Superposition of two HI phases}

A more realistic case would be to treat the HI as consisting of two phases or
components, characterized by a warm ($v_{z}$ = 11 $kms^{-1}$) and a cold medium
($v_{z}$= 7 $kms^{-1}$) respectively. These values match the range seen in the above
fits and represent the two phases as observed in the Galaxy (Kulkarni \&
Heiles 1988). However,observationally the fraction of mass in these two
phases as a function of radius is not known. Assuming that this fraction
is constant with radius, we let its value vary as a free parameter.

The best-fit ${\chi}^{2}$ in this case is 13.7 as compared to 2.8 for the case
with a velocity gradient treated earlier. Although we do not get as good
a fit as was obtained in the case where there is a gradient in the
velocity, the best-fit core radius $R_{c}$ still comes out to be 2.5 kpc
which is again of the order of $R_{D}$. That the dark matter dominates at
small radii therefore still remains a robust result irrespective of the
input gas parameters used. The best-fit case gives the fraction of HI in
the cold medium to be 0.2.\\

We had taken this ratio to be constant for simplicity. Interestingly, this assumption is justified by the recent detailed study by Dickey et al. (2009) involving absorption and emission spectra in $21$ cm in the outer Galaxy. They use this to map the distribution of the cold and warm phases of the HI medium, and surprisingly find this ratio to be a robust quantity in the radial range of $R_{sun}$ to $3$ $R_{sun}$. They find this ratio is $\sim 0.15 - 0.2$, which agrees well with the best-fit ratio 0.2 that we obtain. It is interesting that this ratio obtained by two different techniques is similar in the two galaxies.

The case with a gradient with a higher velocity dispersion within the optical radius gives the lowest ${\chi}^{2}$ value 
(Fig.2), which we adopt as our best case. We note that this choice is not inconsistent with the constant phase ratio measured by Dickey et al. (2009) which was for the outer Galaxy.

$\bullet$ \textbf{High value of the gas velocity dispersion} 

The high gas velocity dispersion required to get an improved fit to the
scaleheight data is surprising given the superthin nature of the galaxy,
whose small stellar scale height implies that it is among the dynamically
coldest of galactic disks (e.g., Matthews 2000).

The origin of this high gas velocity is beyond the scope of this paper.
However, independent of its origin, this high value of the gas velocity
dispersion can partly explain why star formation is inefficent in
UGC 7321. This is because, to first order, a higher gas dispersion will
tend to suppress star formation since Toomre Q criterion ( Q $<$ 1) is
less likely to be satisfied, hence the disc is less likely to be unstable
to star formation. \\

\section{Conclusions}

We have modelled the LSB superthin galaxy UGC 7321 as a gravitationally-coupled system of stars and {\mbox{H\,{\sc i}}} gas, responding to the gravitational potential of the dark-matter halo, and used the observed rotation curve and the {\mbox{H\,{\sc i}}} vertical scale heights as simultaneous constraints to determine the dark halo parameters. We find that the best-fit gives a spherical, isothermal halo with a central density in the range of 0.039-0.057 $M$$_{\odot}$ $pc$$^{-3}$ and core radius of 2.5-2.9 $kpc$. The value of the best-fit core density is comparable to values obtained for HSB galaxies. The core radius is comparable to that of the disc scale length unlike HSB galaxies studied by this method, implying the importance of the dark-matter halo at small radii in UGC 7321. Thus we find that UGC 7321 is dark matter dominated at all radii, and the results of our analysis support the idea that the thinnest of the galaxies are the most dark matter dominated.\\

\end{document}